\documentclass{article}

\usepackage{geometry}
\def\thebibliography#1{
 \list
 {[\arabic{enumi}]}{\settowidth\labelwidth{[#1]}\leftmargin\labelwidth
 \advance\leftmargin\labelsep
 \usecounter{enumi}}
 \def\newblock{\hskip .11em plus .33em minus -.07em}
 \sloppy
 \sfcode`\.=1000\relax}
 
\usepackage{bibentry}

\usepackage{graphicx}
\usepackage{empheq}

\usepackage{natbib}
\usepackage{multirow}
\usepackage{amsmath,amssymb}
\usepackage{wasysym}

\usepackage{color}

\begin{document}

\markboth{X. Yu et al.}
{PIV of swirling flow in a conical pipe with vibrating wall}


\title{PIV of swirling flow in a conical pipe with vibrating wall}


\author{Yan Xu
\\
Department of Mechanical Science and Engineering, 
Northeast Petroleum University, 
\\
Daqing, China,
\\
SFMG, Department of Mechanical Engineering, 
The University of
\\
Sheffield, Sheffield, UK
\\[2ex]
Yan-yue Zhang
\\
Department of Mechanical Science and Engineering, Northeast Petroleum University, 
\\
Daqing, China
\\[2ex]
F.C.G.A Nicolleau
\\
SFMG, Department of Mechanical Engineering, The University of Sheffield,
\\
Mappin Street S1 3JD Sheffield, UK
\\[2ex]
Zun-ce Wang
\\
Department of Mechanical Science and Engineering, Northeast Petroleum University, 
\\
Daqing, China}

\maketitle


\date{\today}


\begin{abstract}
Swirling flows in conical pipe can be found in a number of industrial processes, such as hydrocylone, separator and rotating machinery. It has been found that wall oscillations can reduce the drag in water channel and pipe flows, but there is no study of a swirling flow combined with a vibrating wall in conical pipes, though there are many applications of such combination in engineering processes. A two-dimensional particle image velocimetry (PIV) is used to measure the swirling flow field in a water conical pipe subjected to a periodic vibrating wall for a Reynolds number 3800. The flow velocity statistics are studied under different vibration frequencies corresponding to Strouhal numbers from 60 to 242. The instantaneous axial and vertical velocity in one vibrating period, the mean velocities, and Reynolds stresses were obtained. The results show the existence of an intermediary recirculation cell in the middle of the pipe. They also show that the vibration improves the symmetry for the swirling flow while decreasing dramatically the velocity fluctuation.

\end{abstract}


\maketitle

\section{Introduction}

Swirling flows in conical pipe can be found in a number of industrial processes, such as hydrocylone, separator and rotating machinery. 
Drag reduction in turbulent flows has been the focus of recent research on fluid mechanics. Researchers have found that wall oscillations can reduce the drag in water channel and pipe flows. When the wall oscillation is optimized with the appropriate amplitude and frequency, a significant drag reduction can be achieved by as much as 40\%. 
A swirling flow can make the pressure drop higher but it is important in hydrocyclones and separators to reduce the pressure drop while retaining high rotating velocities in order to improve a separator efficiency. The current work aims to investigate the drag reduction of swirling flows in a conical pipe.

The performance of the swirling flow has been extensively investigated over the past decades. We just give a brief literature review here.
Hydrocyclones have been comprehensively studied numerically by 
\cite{Nowakowski2008}, \cite{Doby2005} and \cite{Doby2007}. 
These studies highlighted the basic mechanisms of a swirling flow.
Flow patterns were investigated by 
\cite{Hsieh-Rajamani-1991} in a 75~mm hydrocyclone separator. 
They demonstrated the
importance of the turbulent mixing in the feed section as a result of shock effects.
\cite{Lim-et-al-2010} studied the flow patterns and characterised the pressure and velocity distributions 
within a hydrocyclone separator system experimentally using PIV and computationally using LES. 
\cite{Jakubowski-2015} proposed secondary flow identification methods and identified the secondary flows occurring in a whirlpool using PIV experiments.
\cite{Marins-et-al-2010} investigated the velocity fluctuations of the three-dimensional flow occuring in a hydrocyclone without an air-core using LDA and PIV techniques. 
\cite{Shi-et-al-2012} studied a new vane-type pipe separator (VTPS) by using PIV and measured the swirling velocity field distribution. 
They characterised the tangential and axial velocity profiles.
%

\cite{Cui-et-al-2014}
studied experimentally the flow field in a hydrocyclone separator system ($\diameter $ 50 mm) with air-core  using PIV and computationally using Reynolds  stress  model  (RSM). Vortices and axial velocity distributions obtained by  numerical and experimental methods were in good agreement.
\cite{Oruc-et-al-2016} used particle image velocimetry (PIV) to investigate the passive control of water flow downstream of a circular cylinder ($\diameter$ 50 mm). Their study confirmed that the flow can be successfully controlled by removing large-scale vortical flow structures. 

Reynolds shear stress and turbulent kinetic energy distributions related to the fluctuating velocity components measured in the wake region and
flow-induced vibrations in pipes have been actively studied for a long time. 
\cite{Ibrahim-2010,Ibrahim-2011} highlighted that a pipe conveying fluid may exhibit a variety of dynamic behaviours.
\cite{Zhang-2013,Zhang-et-al-2014}
studied the influence of external excitations on a pipe flow. 
\cite{Zhou-et-al-2017}
investigated chaotic vibration motions in pipes conveying pulsating fluid. The analysis indicates that under small perturbations, complicated dynamics may occur.
The role of spanwise wall oscillations has been found to be very important in drag reduction. 

The influence of the spanwise wall oscillations on turbulence in channel flows has been studied by means of numerical and experimental studies. It was first studied by \cite{Jung-et-al-1992} and \cite{Akhavan-et-al-1993} who found a continuous reduction of drag and turbulence intensity.

%

\cite{Choi-et-al-2002}, \cite{Huang-et-al-2004}
studied numerically the drag reduction in channel and pipe flows subjected to spanwise wall oscillations. 
See also \cite{Yu-et-al-2014} for the effect on heat transfer.
\cite{Quadrio-Ricco-2004} carried out DNS studies on the turbulent wall shear stress in a turbulent channel flow forced by lateral sinusoidal oscillations of the walls. 
Researches on pipe flow oscillating around its axis conducted by \cite{Quadrio-Sibilla-2000} showed also a drag reduction for this geometry.
\cite{Touber-Leschziner-2012} used DNS to study a fully developed channel flow subjected to oscillatory spanwise wall motion and discussed the fundamental mechanisms responsible for the reduction in turbulent friction drag resulting from the spanwise wall motion. 
Results show that an oscillation period $T u_{\tau}^2/ \nu $ of the order of $10^2$ leads to a strong decline in the level of the Reynolds stresses.
\cite{Ge-Jin-2017}
carried out DNS studies on the turbulent channel flow and showed that spanwise wall oscillation can lead to attenuation of the turbulent enstrophy, which directly enhances turbulent dissipation leading to lower-drag conditions.
\\[2ex]
From this brief review, it appears that 
wall oscillations can reduce the drag in water channel and pipe flows, and researchers have used PIV to study the swirling flow field and to measure the effect of vibration on the flow field. 
But to our knowledge there is no study of a swirling flow combined with a vibrating wall in conical pipe, though there are many applications of such combination in engineering processes - such as hydrocyclone to name one. In this paper, PIV is used to measure the swirling flow field in a conical pipe subjected to a  periodic axial vibration. 
\\[2ex]
The paper is organised as follows: in \S~\ref{method} we describe the experimental set up,\ its notations and its
parameters. 
The results are then presented and discussed in \S\ref{secresults}.
The main conclusions are summarised in  \S~\ref{seconcl}.

\section{Experimental set up}
\label{method}

\begin{figure}[htb]
\begin{center}
\includegraphics[width=\textwidth]{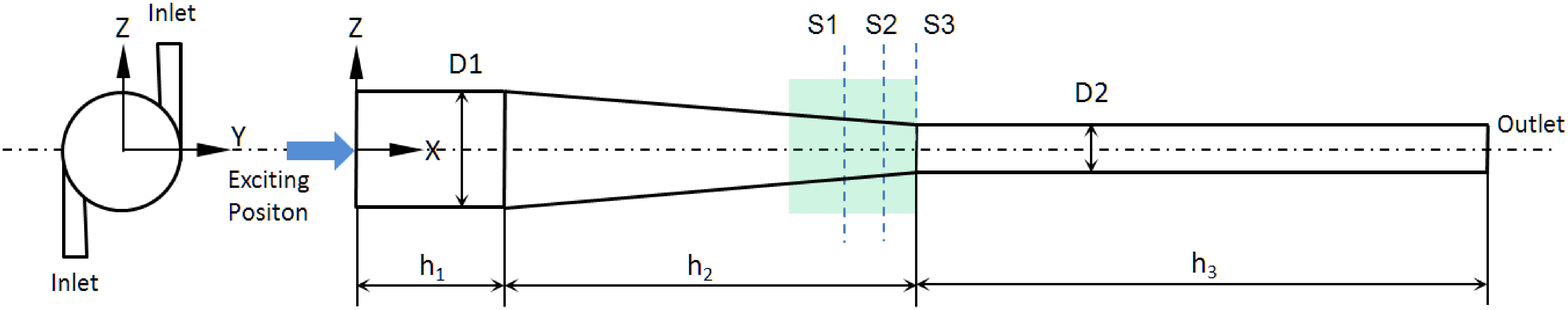}
\includegraphics[width=0.32\textwidth]{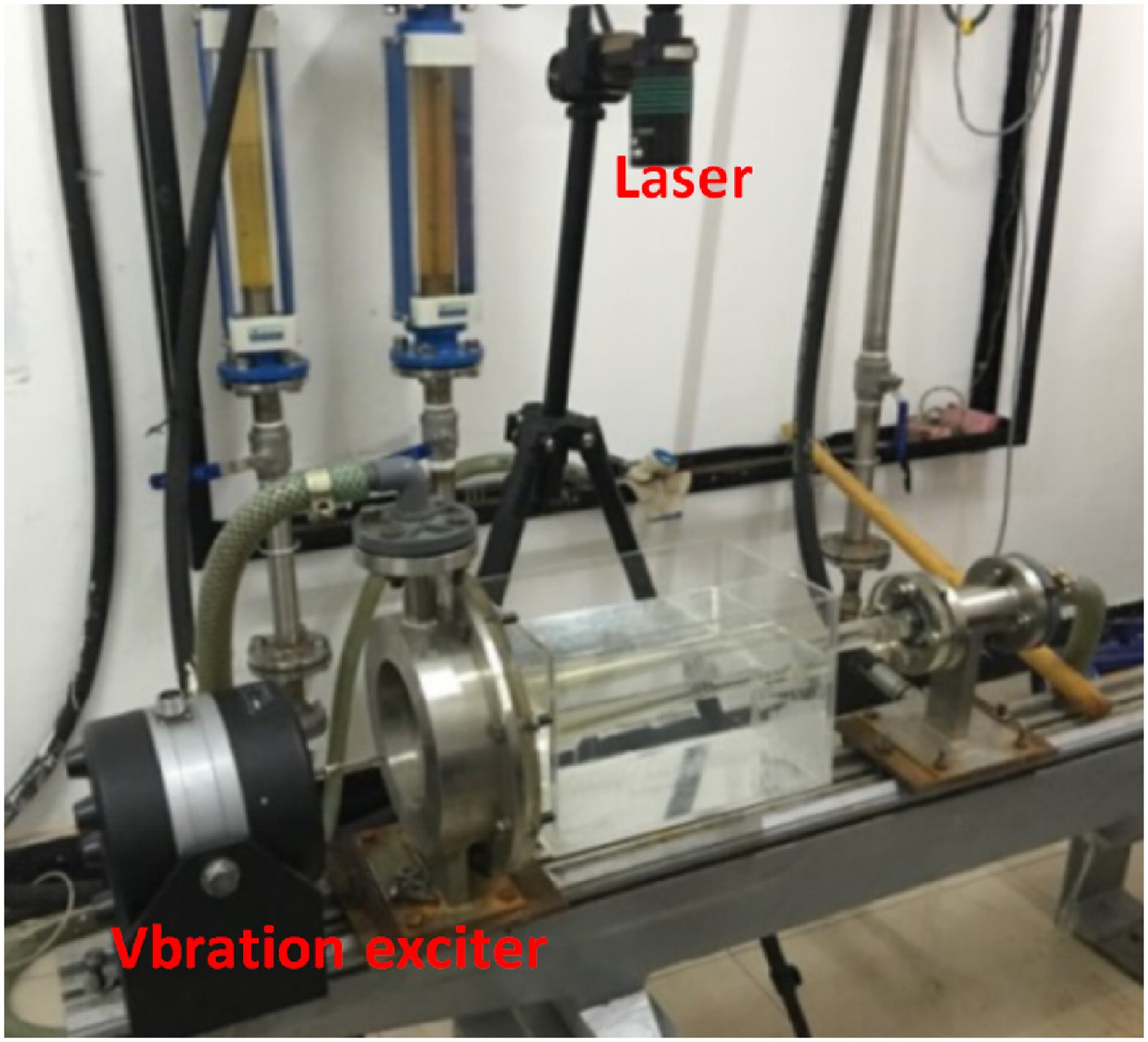}
\includegraphics[width=0.32\textwidth]{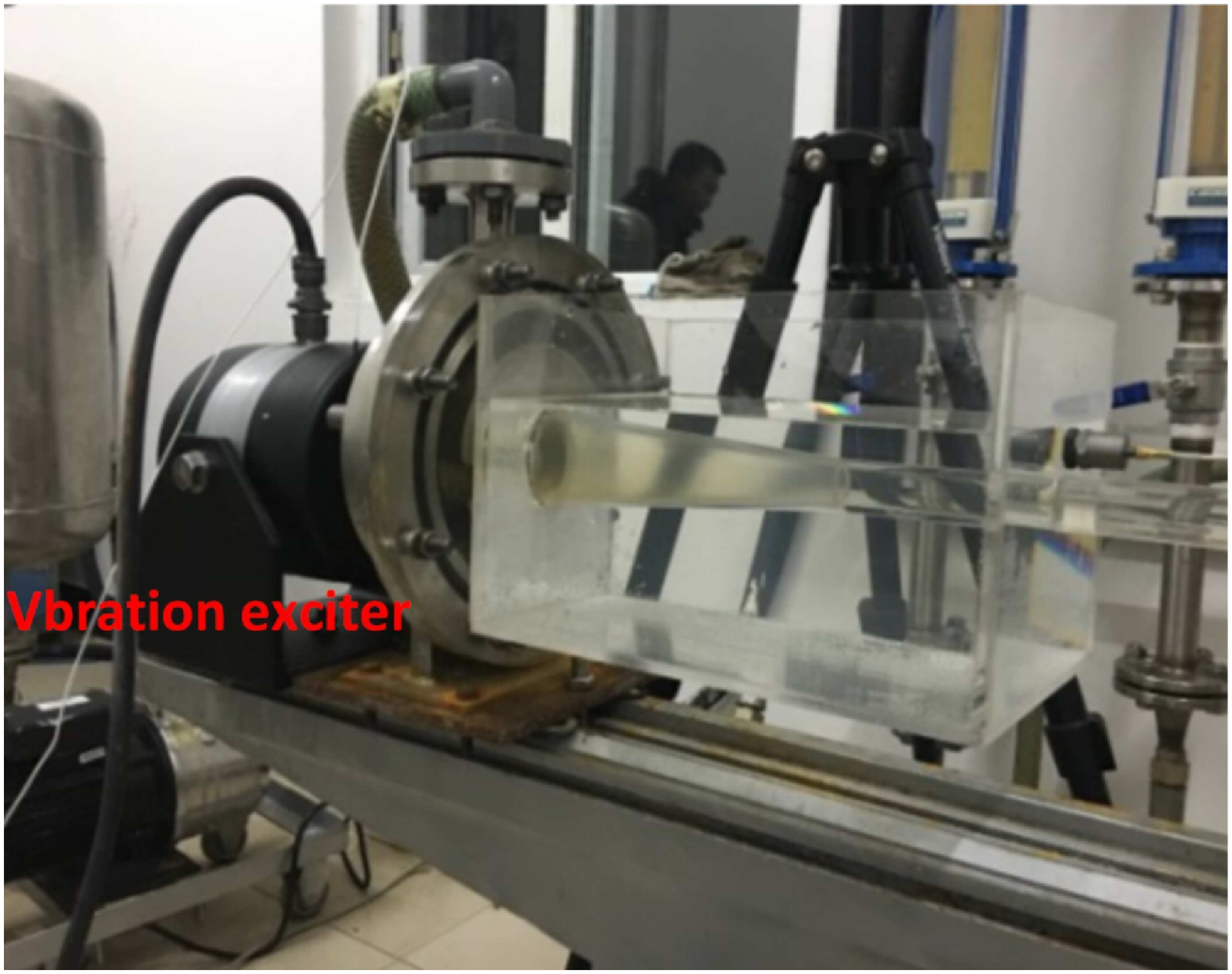}
\includegraphics[width=0.32\textwidth]{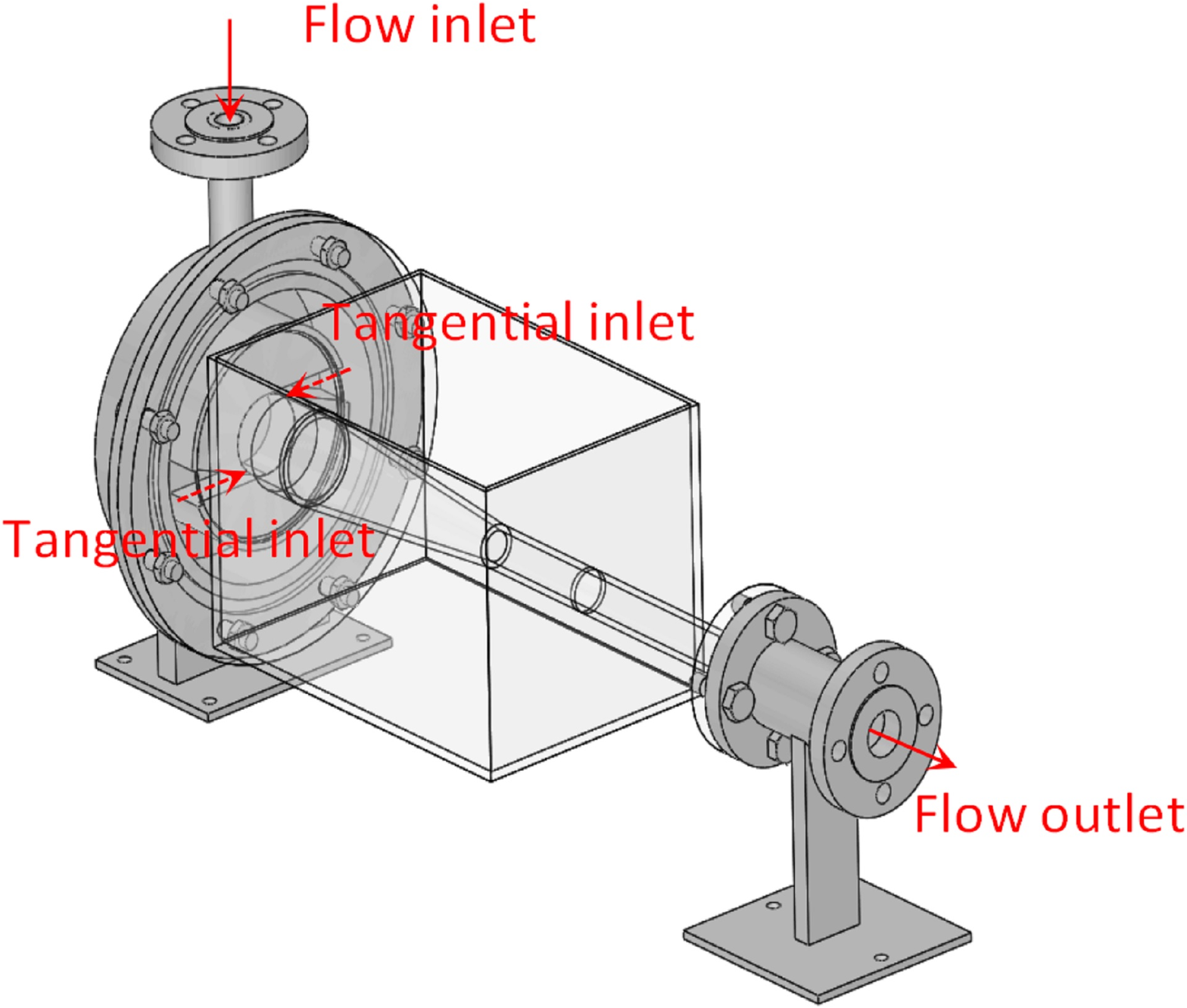}
\end{center}
\caption{\label{Figure1-setup}
Experimental set up, conical pipe geometry,
details of the inlet injection at the bottom}
\end{figure}

The axi-symmetrical pipe geometry is shown in Fig.~\ref{Figure1-setup}. It consists of three pipe elements. The pipe entry is a cylinder of diameter $D_1$ and length $h_1$, the pipe exit a cylinder of diameter $D_2$ and length $h_3$.
The working section is a cone linking these two cylinders. Its length is $h_2$.

The flow is injected in the first pipe element as shown in Fig.~\ref{Figure1-setup} (bottom)
with two opposite inlets in the vertical direction ($z$). 
The velocity at these inlet ist $U_{in}$.
The swirling flow is  generated by the two tangential flow injections at the inlet it then enters the cylindrical chamber that is immediately followed by the conical section.
The flow exits at the extremity of the pipe element of diameter $D_2$. 
Deionized water is used as the working fluid.
The water pumped from the water tank using a centrifugal pump is fed into the conical pipe under controlled flow condition by a set of valves. The flow rate is 0.6~m$^{3}$~h$^{-1}$, the pipe inlet mean velocity, $U_{mean}$, 0.068~m~s$^{-1}$, the Reynolds number at the cone inlet is $Re_{1}= 3800$.
The pipe geometry and flow properties are summarised in Table~\ref{tab1}.
Before taking measurements, in order to ensure that the flow is steady, it was allowed to settle for about two minutes that is about 30 cycles as the flow rate is 0.6 m$^3$~h$^{-1}$.
\begin{table}[h]
\caption{Experimental parameters.}
{\begin{tabular}{llll}
\hline
quantity & symbol & value
\\
\hline
fluid & &water \\ \\
mean velocity at the inlets & $U_{in}$ & 0.135~m~s$^{-1}$
\\
\\
cone inlet diameter &$D_1$ & 56 mm \\
cone outlet diameter &$D_2$ & 21 mm & 0.375 $D_1$
\\ \\

mean velocity at $D_1$ & $U_{mean}$ & 0.068 m~s$^{-1}$
\\
inlet Reynolds number at $D_1$ & $Re_{1}$ &   3800
\\
volume flow rate & $\dot{Q}$ & 0.6~m$^3$~h$^{-1}$ & 0.167~$\ell$~s$^{-1}$
\\
\\
inlet pipe length & $h_1$ & 83 mm  & 1.482 $D_1$
\\
conical pipe length & $h_2$ & 200 mm & 3.571 $D_1$
\\
outlet pipe length & $h_3$ & 300 mm & 5.357 $D_1$
\\
\\
1st profile distance from inlet & $S_1$ & 263~mm     & 4.696 $D_1$
\\
2nd profile distance from inlet & $S_2$ & 273~mm     & 4.875 $D_1$
\\  
3rd profile distance from inlet & $S_3$ & 283~mm     & 5.053 $D_1$
\\
\\
excitation frequencies & $f$ &  50~Hz, 100~Hz, 200~Hz
\\
Strouhal number & $St$ & 60.7, 121, 242
\\
\hline
\end{tabular}
\label{tab1}} 
\end{table}

\begin{figure*}[htb]
\begin{center}
\includegraphics[width=0.5\textwidth]{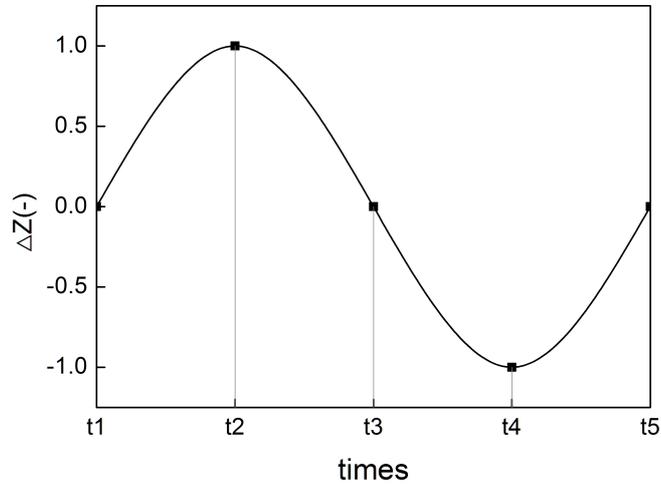}
\end{center}
\caption{\label{Fig7-sinu}
Sinusoidal vibration showing the reference times $t_1$, $t_2$, $t_3$ abd $t_4$. }
\end{figure*}

A sinusoidal vibration (Fig.~\ref{Fig7-sinu}) is then imposed on the pipe wall at the inlet in the axial direction $x$.
We use $f=50$~Hz, 100~Hz and 200~Hz as excitation frequencies. Or in non-dimensional form $St= 60.7$, 121 and 242, where we define $St$ the Strouhal number as
\begin{equation}
St = \frac{f \, U_{mean}}{D_1}
\end{equation}
The vibration amplitude is $1$~mm.
The oscillation generating system consists of a frequency converter (HEA-200C), exciter (HEV-200). Its main function is to provide harmonic vibrations at different frequencies for the test section.

\subsection{Facilities, PIV system}

\begin{figure*}[htb]
\begin{center}
\includegraphics[width=0.8\textwidth]{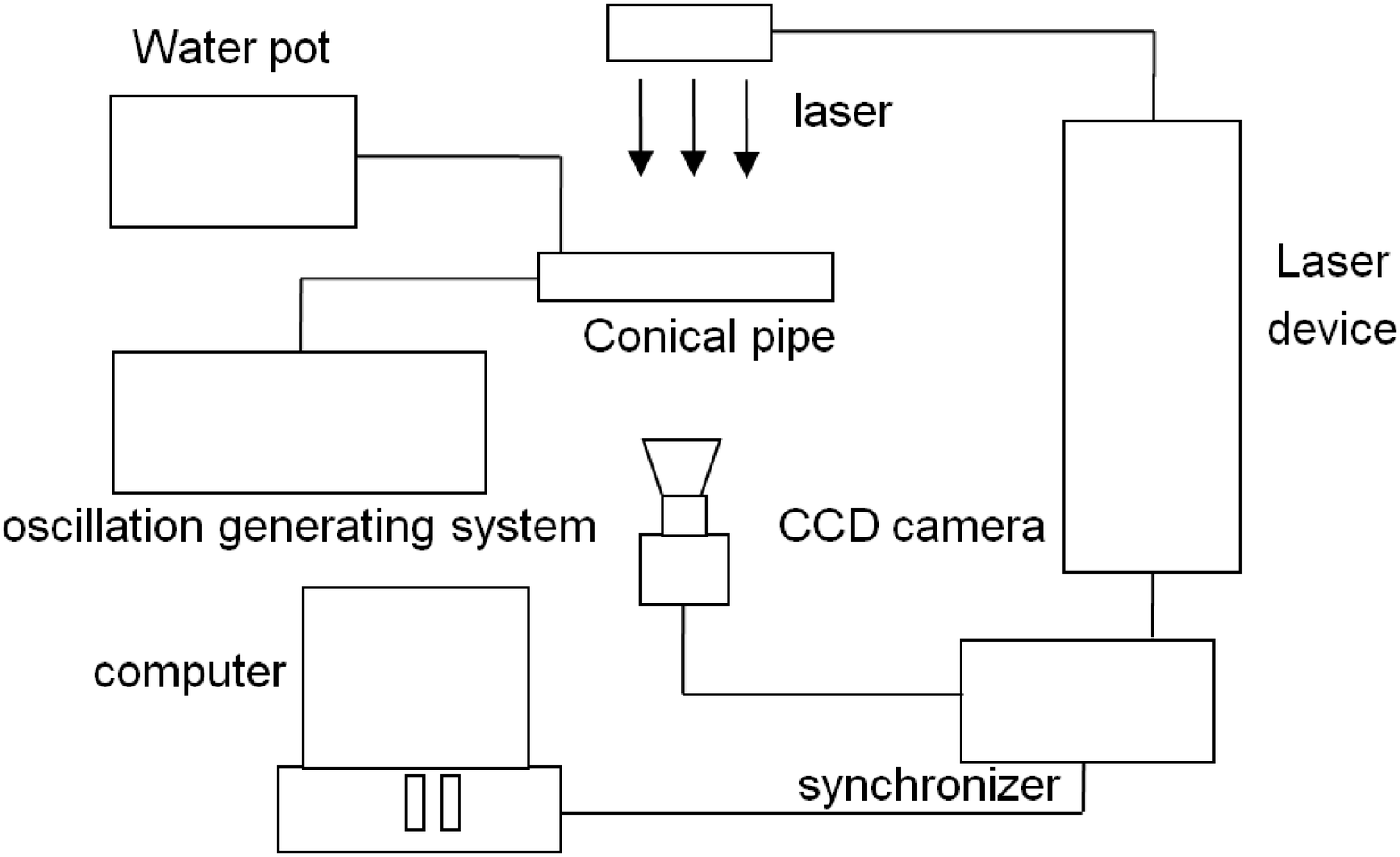}
\end{center}
\caption{\label{Figure2-PIVsetup}
PIV setup}
\end{figure*}

The schematic diagram of the experimental apparatus is illustrated in Fig.~\ref{Figure2-PIVsetup}. 
%
The measurements were carried out in the conical pipe element made of plexiglas.
The deionized water was seeded with polystyrene particle tracers 5~$\mu$m in size. 
The flow velocity is measured using PIV in the greyed vertical plan shown in Fig.~\ref{Figure1-setup} that is in the vertical plane of axi-symmetry $XOZ$ ($y=0$). In particular, velocity profiles are presented at $x=S_1$, $S_2$ and $S_3$ - see Fig.~\ref{Figure1-setup}. $S_3$ corresponds to the conical section end.
\\[2ex]
The PIV parameters are summarized in Table~\ref{tab2}.
\begin{table}[h]
\caption{PIV parameters.}
{\begin{tabular}{ll}
\hline
light sheet thickness $\Delta_0$ & 1.5~mm
\\
seeding particle diameter & 5~$\mu$m
\\
pulse frequency & 10$^4$~Hz
\\
\hline
\end{tabular}
\label{tab2} }
\end{table}

The 2D-PIV measurements were performed with a Dantec Dynamics A/S system. The light source was produced by a LPU550 solid pulse laser that produced high energy (200~mJ) pulses of green light (532~nm). The collimated laser beam was transmitted through a lens to generate a $\Delta_0=1.5$~mm thick lightsheet. The reflected light was recorded at a frequency of 14~Hz by the CCD camera (FlowSence EO 4M) with $2048 \times 2048$~pixels resolution. 
Images were calibrated from pictures of a reference target specially designed for the present purpose. 

The water was seeded with polystyrene particle tracers 5 $\mu$m in size gradually. We counted the number of particles in the interrogation window until it typically contains 6 to 10 particles per image.

The velocity in the conical pipe at an instant between each two consecutively recorded images was obtained by mean of PIV. Adaptive correlation (Dynamic Studio software) was processed on $32\times32$ pixels-size final interrogation spots, Fast Fourier Transformation was adopted, with 50\% overlap, which gives a $32 \times 32$ grid step size. 

Validation is performed by first applying peak validation on the image cross-correlation which gives a peak height of 0.1 and a peak height ratio of 1.2, and secondly by comparing each vector to its neighbours using the universal outlier detection algorithm. Vectors that deviate too much from their neighbours can be replaced by the average of the neighbours as a reasonable estimate of true velocities. Average Filter is used to filter out vector maps by arithmetically averaging over vector neighbours. No further filtering has been applied to the velocity fields in order to keep the whole measurement information. 

During the experiment, to minimize the effects of reflection and refraction of the light beams, 
the conical working element was enclosed in a water-filled perspex cube tank of square section. 
In order to satisfy the high-image-density criterion, the interrogation window typically contained 6 to 10 particles per image. 

For each experimental condition, the CCD camera captures 200 instantaneous velocity vector fields in the XOZ plan. The data were collected in 15~s. 
In order to achieve data sets large enough and reduce experimental error while not exceeding the camera capacity, the time delay between pulses ranged as 100~$\mu$s (Cross frame time). According to the moving distance of the image particles in the shooting, in this experiment, the smallest cell size is $32\times 32$. According to the flow velocity, the inter frame time is guaranteed to be 1/2 of the length of the particle moving in the smallest cell.

\subsection{Experimental data analysis method}
\label{sec2.2}

With PIV we can measure two components $(U,W)$ of the instantaneous Eulerian velocity at each point in the laser sheet.
The instantaneous velocity at a point $(x,z)$ and time $t_n$ obtained by PIV is $(U(t_n),W(t_n))$. $t_n$ is the nth measurement time.
The time average velocities at point $(x,z)$  are
\begin{align}
\overline{U} = \frac{1}{N} \sum_{n=1}^{N} U(t_n)
\\
\overline{W} = \frac{1}{N} \sum_{n=1}^{N} W(t_n)
\end{align}
Where N is the total number of time samples. 
So the fluctuating velocities $(u(t_n),w(t_n))$ are defined as
\begin{align}
u(t_n) &= U(t_n)- \overline{U}
\\
w(t_n) &= W(t_n)- \overline{W}
\end{align}
The instantaneous Reynolds stresses are given by
$u_i(t_n)u_j(t_n)$ where $i,j=1,3$ and the average Reynolds stresses are defined as
\begin{align}
\overline{u u} = \frac{1}{N} \sum_{n=1}^{N} u(t_n) u(t_n)
\\
\overline{w w} = \frac{1}{N} \sum_{n=1}^{N} w(t_n) w(t_n)
\\
\overline{u w} = \frac{1}{N} \sum_{n=1}^{N} u(t_n) w(t_n)
\end{align}
For the sake of comparison we define a mean velocity based on the diameter $D_1$ as
\begin{equation}
U_{mean} = \frac{4 \dot{Q}}{\pi {D_1}^2} = 0.068 \text{~m~s}^{-1}
\end{equation}

\section{Results}
\label{secresults}

\subsection{Instantaneous velocity in one vibrating period - $S_2$-section}

\begin{figure*}[h!]
\begin{center}
\includegraphics[width=0.49\textwidth]{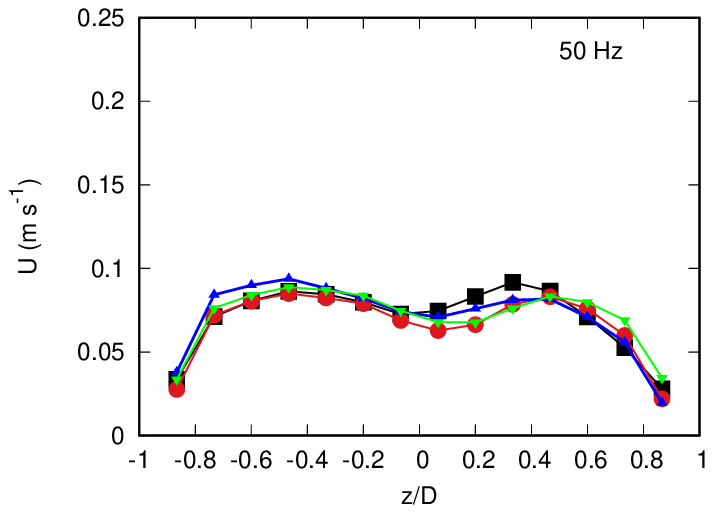}
\includegraphics[width=0.49\textwidth]{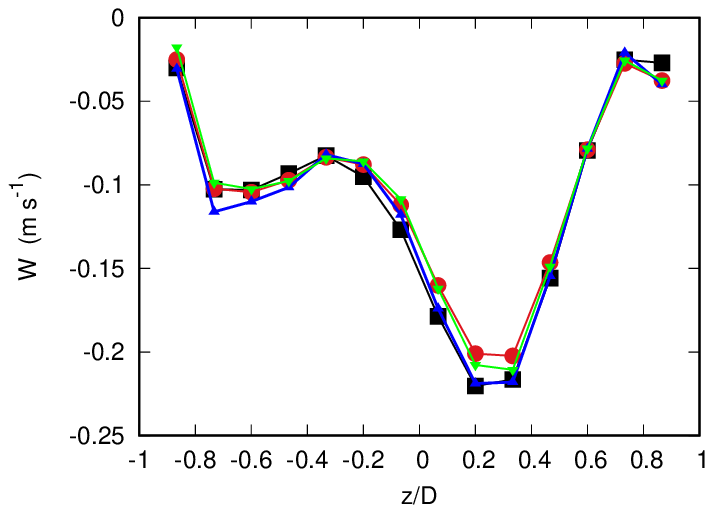}
\includegraphics[width=0.49\textwidth]{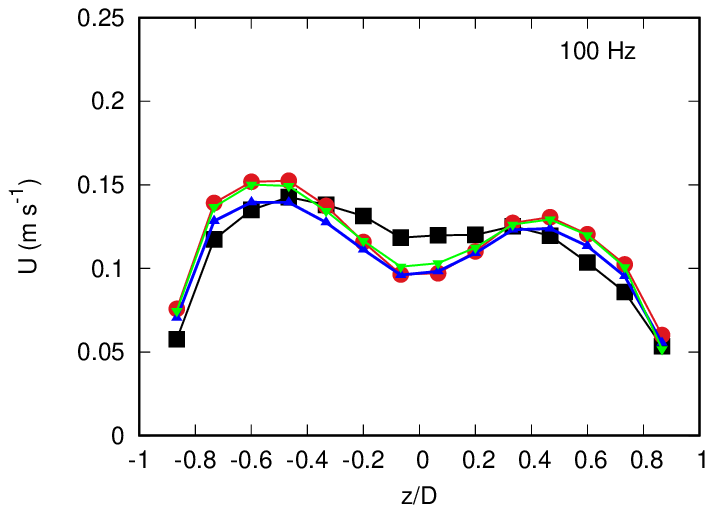}
\includegraphics[width=0.49\textwidth]{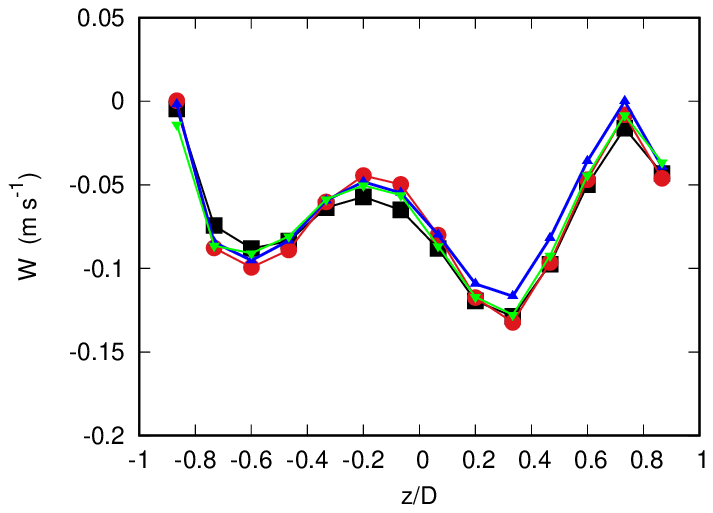}
\includegraphics[width=0.49\textwidth]{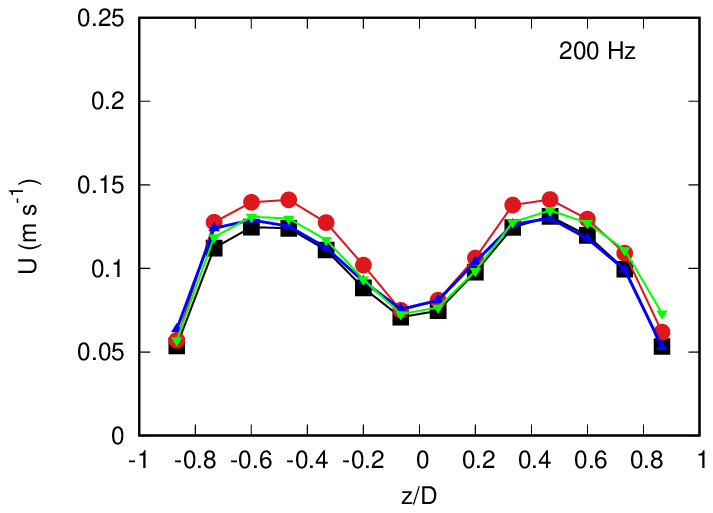}
\includegraphics[width=0.49\textwidth]{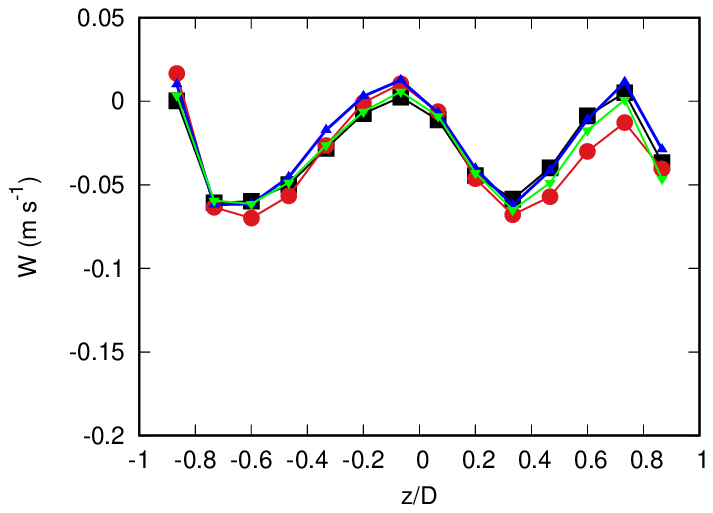}
\end{center}
\caption{\label{Figure5} Instantaneous velocity profiles at $\blacksquare$ $t_1$, \textcolor{red}{\textbullet}
$t_2$,  \textcolor{blue}{$\blacktriangle$} $t_3$ and \textcolor{green}{$\blacktriangledown$} $t_4$ in section 2; top row to bottom row: 50~Hz, 100~Hz and 200~Hz.
Left column axial velocity $U$, right column vertical velocity $W$.}
\end{figure*}

In Fig.~\ref{Figure5}, we plot the instantaneous profiles at 4 characteristic times in the vibration period. The different
times are shown in Fig.~\ref{Fig7-sinu}, they subdivide the signal in 4 equal intervals.
\\
For these instantaneous profiles, there is little effect of the time at which the profile is measured.
The vertical velocity is always negative (we will come back to this point later on). As the frequency increases some oscillations in the profile can be seen.


\subsection{Effect of vibration on the mean velocities}

\begin{figure*}[htb]
\begin{center}
\includegraphics[width=0.49\textwidth]{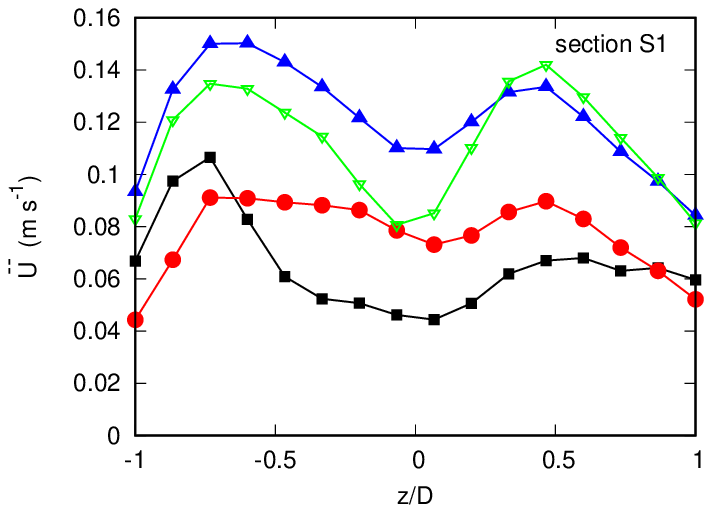}
\includegraphics[width=0.49\textwidth]{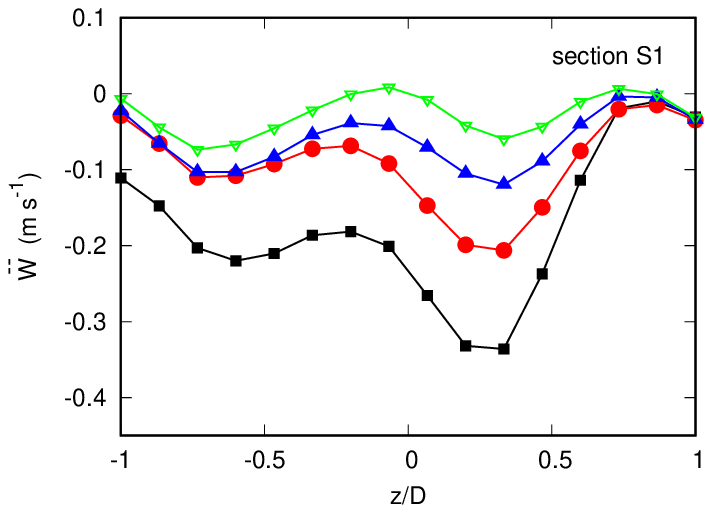}
\includegraphics[width=0.49\textwidth]{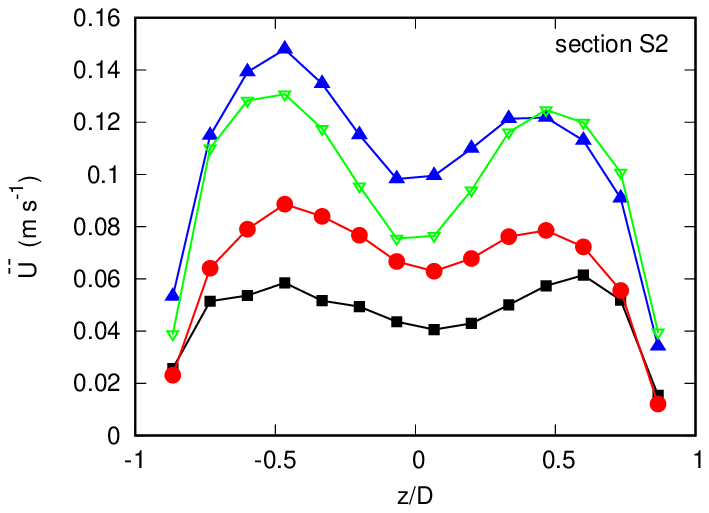}
\includegraphics[width=0.49\textwidth]{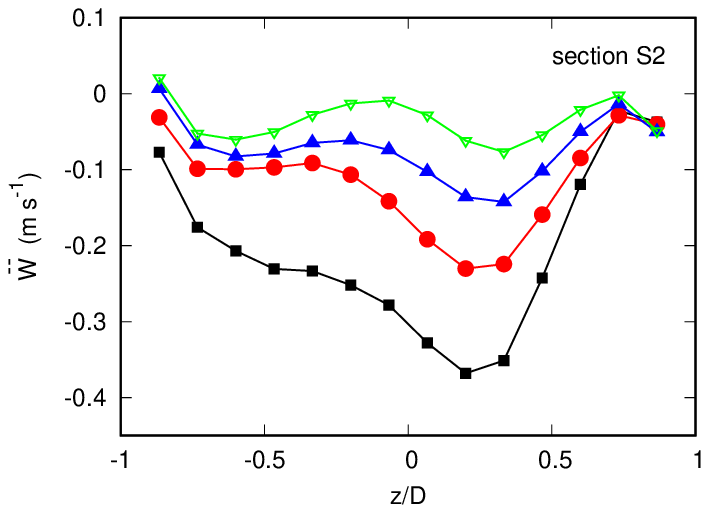}
\includegraphics[width=0.49\textwidth]{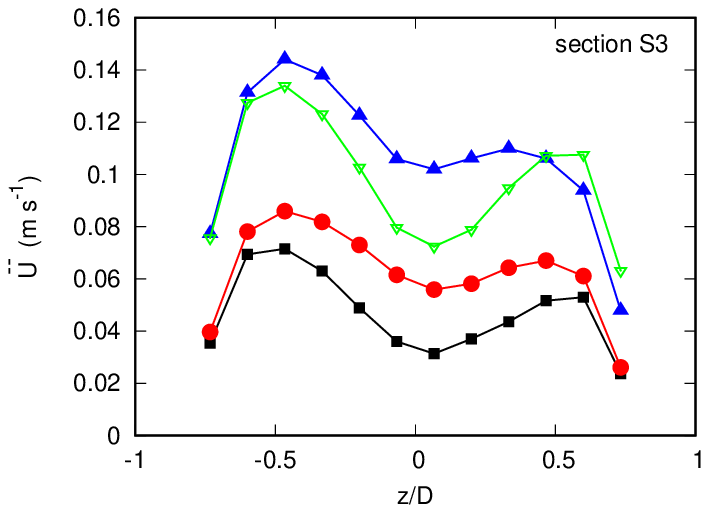}
\includegraphics[width=0.49\textwidth]{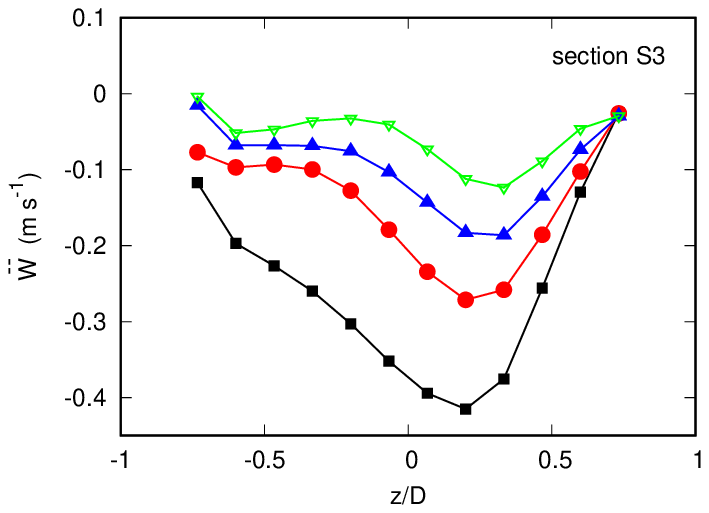}
\end{center}
\caption{\label{Fig-meanvel}
Mean velocity profiles at the different sections, from top to bottom $S_1$, $S_2$ and $S_3$, and for different vibration frequencies:
$\blacksquare$ no vibration, \textcolor{red}{\textbullet}
50~Hz,  \textcolor{blue}{$\blacktriangle$} 100~Hz, \textcolor{green}{$\blacktriangledown$} 200~Hz.
Left column axial velocity $\overline{U}$, right column vertical velocity $\overline{W}$. }
\end{figure*}

We now study the mean and fluctuation components of the profiles as they are defined in \S~\ref{sec2.2}.
We first study the effect of the periodic vibrations on the mean velocities $U$ and $W$ as functions of $z/D$ as presented in Fig.~\ref{Fig-meanvel}.

\subsubsection{Without vibration}

\begin{figure*}[htb]
\begin{center}
\includegraphics[width=0.25\textwidth]{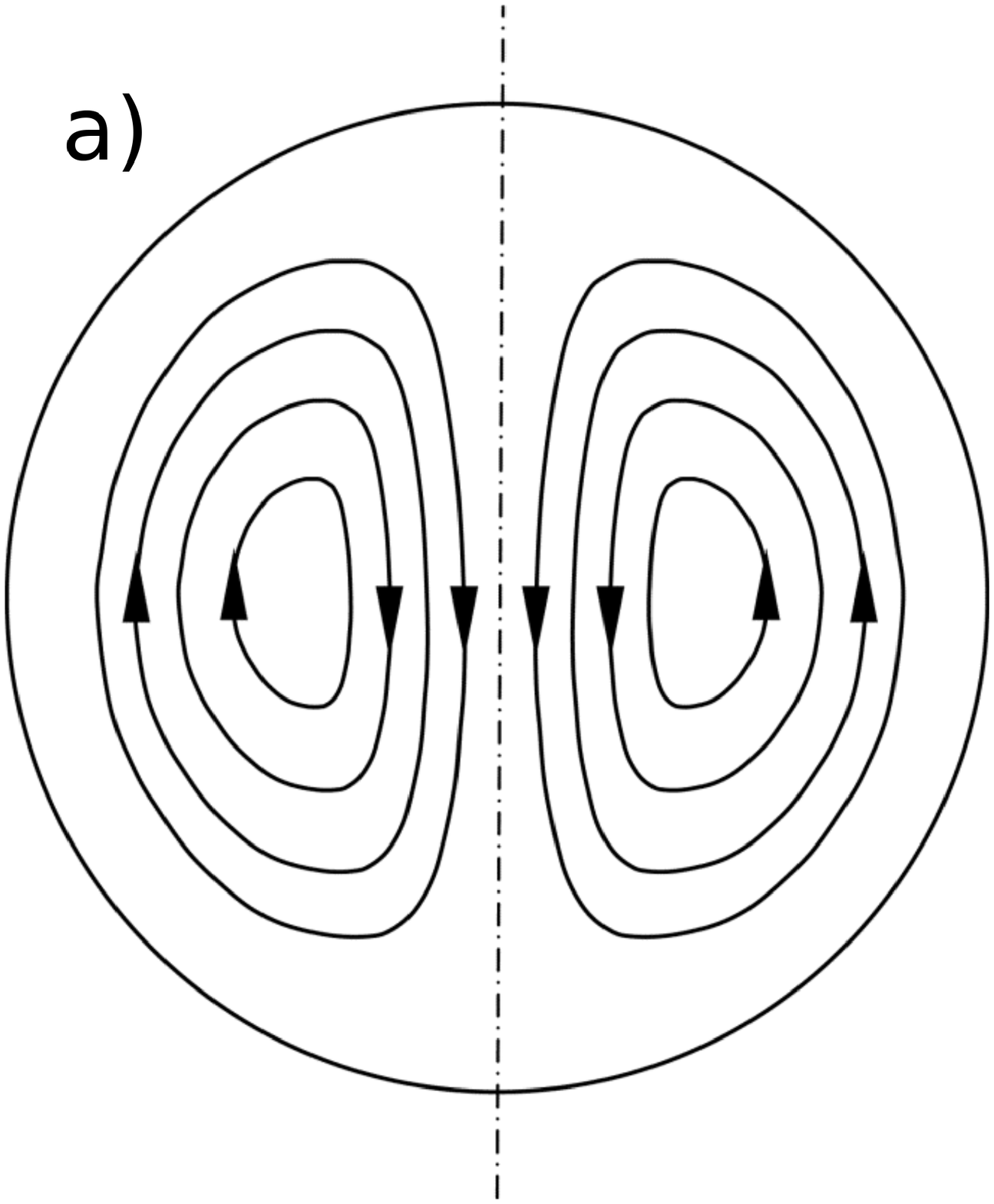} \hspace*{0.5cm}
\includegraphics[width=0.25\textwidth]{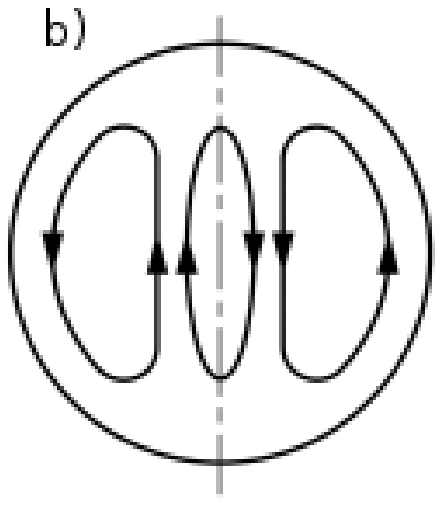} \hspace*{0.3cm}
\includegraphics[width=0.29\textwidth]{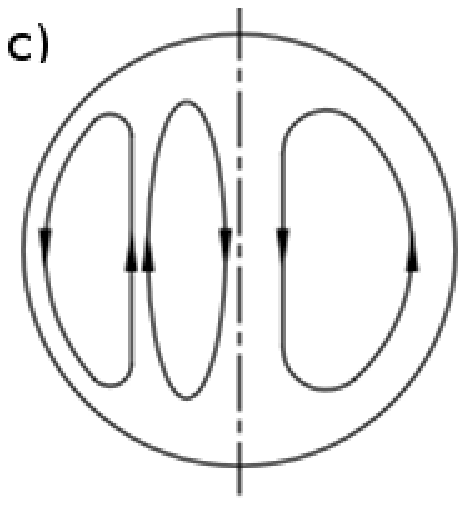}
\end{center}
\caption{\label{Fig-recirc} a) Secondary flow existing after a pipe bend. b) Possible recirculation cells due to tangential velocity. c) A type of recirculation cells showing negative vertical velocity in the laser plan.}
\end{figure*}

In the absence of vibration, except in the first section $S_1$, the axial velocity profile ($\overline{U}$) is close to a symmetric profile and around $U_{mean}$.
The vertical velocities $W$ are all negative indicating the presence of secondary flows formed of at least two recirculation cells of the kind illustrated in Fig.~\ref{Fig-recirc}a. 
This secondary flow is known to exist after a bend in a pipe, here a similar effect would be caused by the forced tangential velocity component at the inlet. However, the
flow topology in Fig.~\ref{Fig-recirc}a is not compatible with a tangential velocity introduced at the inlet. So there must be a odd number of cells as illustrated in Fig.~\ref{Fig-recirc}b.
Furthermore, in order to capture the negative velocity in the laser sheet which is a symmetry plane for the cone, the recirculation cells must not be symmetrical. There must be an even number of cells on one side and an odd number of cells on the other side of the laser sheet as shown in Fig.\ref{Fig-recirc}c.

\subsubsection{With vibration}

The vibrations tend to weaken that secondary cell, as can be observed on the
left column of Fig.~\ref{Fig-meanvel}. As the frequency increases the
vertical velocity decreases in absolute value. This is can be observed in all three sections.

The vibration also increases the meam velocity in the streamwise direction (except at $S_1$ for $f=50$~Hz in the limited range 
$-1 \le z/D_1 \le -0.75$). But this can only be true in this particular vertical laser plan. Indeed the mass flow rate is constant so the streamwise velocity cannot be increased everywhere.
There is no monotonic effect of the frequency on the streamwise mean velocity $\overline{U}$. The maximum increase is obtained for $f=100$~Hz and then the increase is slightly less for $f=200$~Hz.
This is indeed to be expected as the mass flow rate must be conserved.

The vibration effect on the vertical velocity ($\overline{W}$) is more systematic. 
It is monotonically increased in relative value, decreased in absolute value as the frequency increases (except perhaps for $z/D_1 \ge 0.75$ where the effect of vibration is negligible). This could be an indication that the secondary flow is damped and the tangential flow near the wall is strengthened.

\subsection{Reynolds stress on section $S_2$}

\begin{figure*}[htb]
\begin{center}
\includegraphics[width=0.49\textwidth]{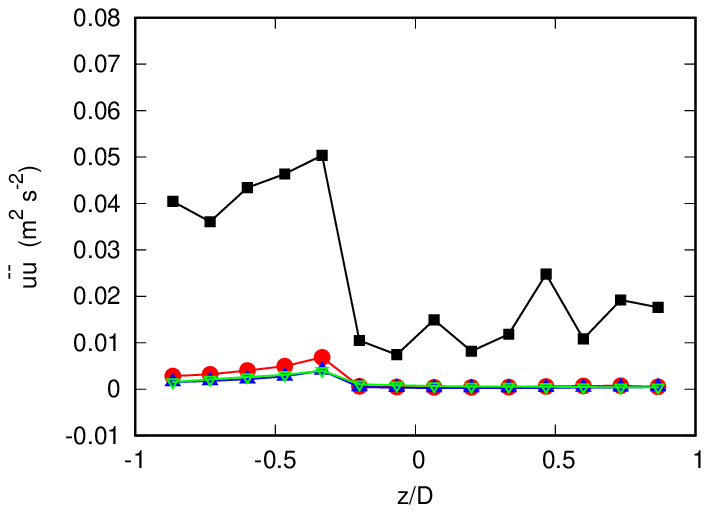}
\includegraphics[width=0.49\textwidth]{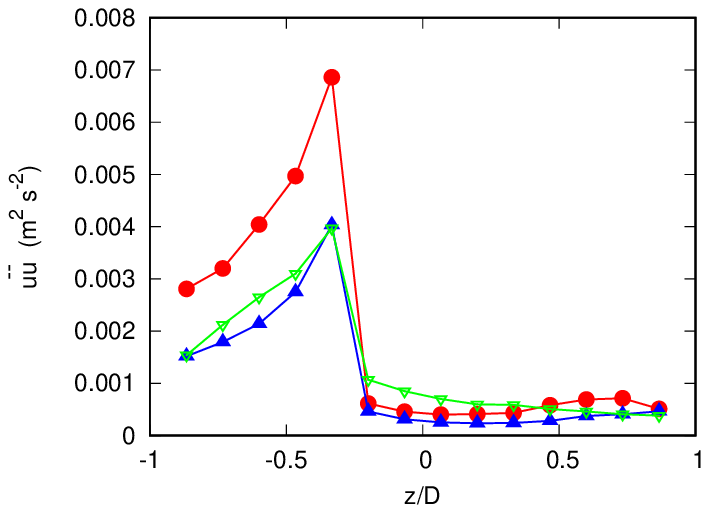}
\includegraphics[width=0.49\textwidth]{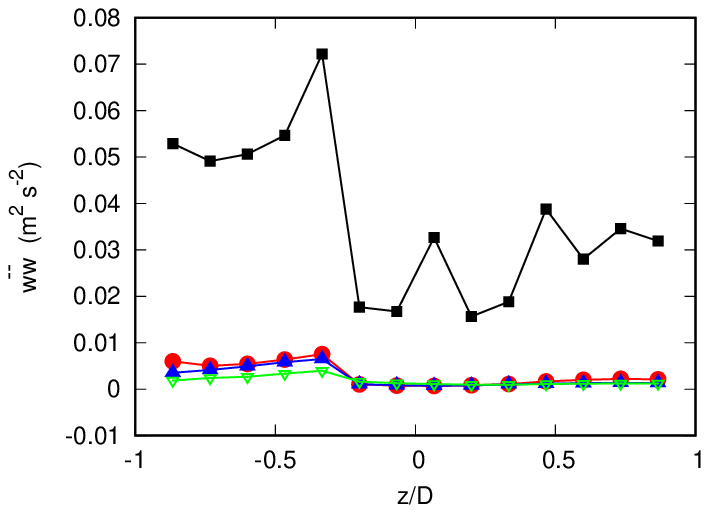}
\includegraphics[width=0.49\textwidth]{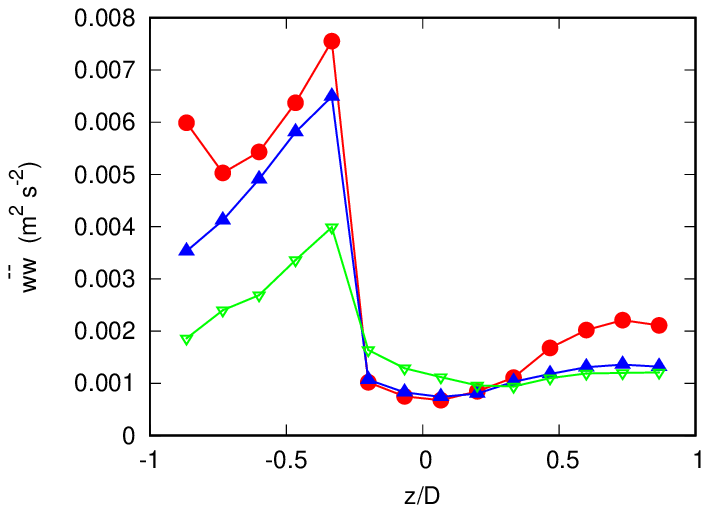}
\includegraphics[width=0.49\textwidth]{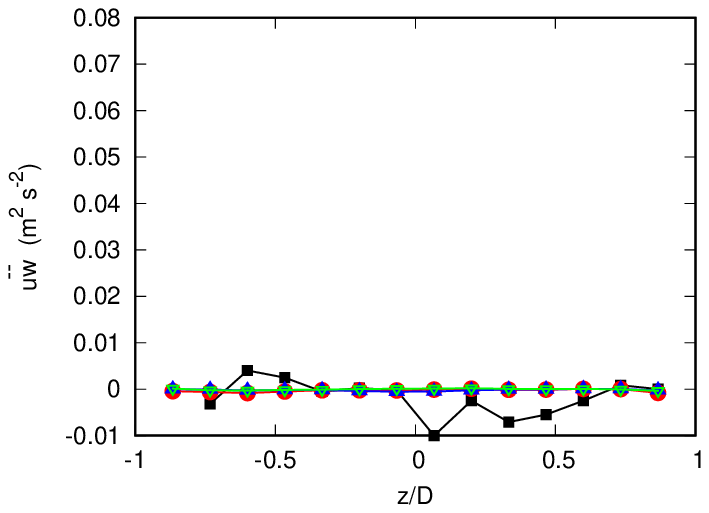}
\includegraphics[width=0.49\textwidth]{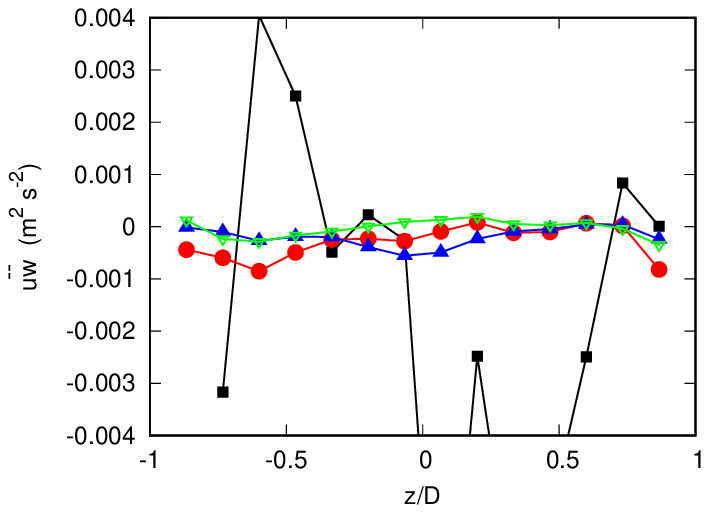}
\end{center}
\caption{\label{Fig-fluc}
Reynolds stresses with different vibration frequencies in section 2,
$\blacksquare$ no vibration, \textcolor{red}{\textbullet}
50~Hz,  \textcolor{blue}{$\blacktriangle$} 100~Hz, \textcolor{green}{$\blacktriangledown$} 200~Hz.}
\end{figure*}

Figure~\ref{Fig-fluc} shows
the effect of the periodic vibrations on the Reynolds stresses. The profiles are presented as functions of the vertical axis ($z/D$) for the three components of the stress tensor $\overline{uu}$, 
$\overline{ww}$ and $\overline{uw}$. 
The vibration clearly decreases the fluctuation intensity. The results for the cases with vibration are emphasised in the left column with a larger magnification. It is clear that the fluctuations drop significantly to about a tenth of their initial value without vibration.
It is worth noting as well that the cross-correlation $\overline{uw}$ is particularly small when vibration is added which is a further indication that the vibrating wall is destroying the scenario conjectured in Fig.~\ref{Fig-recirc}c.

\section{Conclusion}
\label{seconcl}

PIV measurements were conducted in the vertical plane section of a conical pipe. The flow is injected tangentially to create a swirling motion and some harmonic vibrations are applied at the wall in the 
direction of the mean flow.

From the PIV measurements we have clear evidence of a complex flow topology developing in the conical section of the pipe. 
We concluded that there must be an asymmetrical distribution of recirculation zones with a odd number of cells.
That recirculation topology is weakened but not destroyed by the effect of the wall periodic vibrations.
Vibrations have a clear effect on the turbulence fluctuation decreasing them dramatically.
Further work would be necessary in particular in other laser plans as the flow is not axi-symetrical.

\section{Acknowledgement}

This research was supported by the National Natural Science Foundation of China (11402051), the Natural Science Foundation of Heilongjiang Province of China (QC2016003).


\bibliographystyle{wsnatbib}



\end{document}